\documentclass[twocolumn,showpacs,preprintnumbers,amsmath,amssymb,floatfix,nofootinbib]{revtex4-1}
\usepackage{color}   
\usepackage{graphicx}
\usepackage{dcolumn}
\usepackage{bm}
\usepackage{slashed}

\begin{document}

\def\bb    #1{\hbox{\boldmath${#1}$}}

\title{Event-by-Event Study of Space-Time Dynamics in Flux-Tube Fragmentation }

\author{Cheuk-Yin Wong}
\email{wongc@ornl.gov}
\affiliation{Physics Division, 
Oak Ridge National Laboratory, 
Oak Ridge, TN 37831, USA}


\begin{abstract}

In the semi-classical description of the flux-tube fragmentation
process for hadron production and hadronization in high-energy
$e^+e^-$ annihilations and $pp$ collisions, the rapidity-space-time
ordering and the local conservation laws of charge, flavor, and
momentum provide a set of powerful tools that may allow the
reconstruction of the space-time dynamics of quarks and mesons in
exclusive measurements of produced hadrons, on an event-by-event basis.
 We propose procedures to reconstruct the
space-time dynamics from event-by-event exclusive hadron data to
exhibit explicitly the ordered chain of hadrons produced in a flux
tube fragmentation.
As a supplementary tool,  we infer the average space-time coordinates of the $q$-$\bar
q$ pair production vertices 
from the $\pi^-$ rapidity distribution data obtained by the
NA61/SHINE Collaboration
in $pp$ collisions at $\sqrt{s}$ = 6.3 to
17.3 GeV.
\end{abstract}


\maketitle
\section{Introduction}
\label{intro}

The fragmentation of a color flux tube is an important basic process
in the production of low-$p_T$ particles in high-energy $e^+e^-$
annihilations and $pp$ collisions
\cite{Nam70,Bjo73,Cas74,Sch62,Art74,And79,And83,Art84,And83a,Sjo86,Sjo06,Sjo14,Sch51,Wan88,Pav91,Won91a,Won91b,Gat92,Won95,Feo08,Won94}.
In the case of an $e^+e^-$ annihilation, the initial reaction in $e^+ +
e^- \to q + \bar q$ leads to a color flux tube between the quark and
the antiquark that subsequently fragments into produced hadrons.  In
a nucleon-nucleon collision, a quark of one nucleon and the diquark of
the other nucleon form one flux tube, or its idealization as a quantum
chromodynamics (QCD) string.   Subsequent fragmentation of the flux tube leads to the
production of low-$p_T$ hadrons.  It is distinctly different from the
process of relativistic hard-scattering and parton showering in
perturbative QCD that dominate the production of the higher-$p_T$
hadrons \cite{Bla74,Ang78,Fey78,Owe78,Rak13,Eli96,Won12}.

The process of flux-tube fragmentation falls within the realm of
non-perturbative QCD.  A fundamental description of the process from
the basic principles of QCD is still lacking. Simplified field
theoretical descriptions in terms QED2 or QCD2 in one space and one
time coordinates have been presented to understand some gross features
of the process \cite{Cas74,Sch62,Won94,Won91,Abd01,Fri93,Won10,Kos12}.
However, phenomenological applications of the quantum field
theoretical description remain quite limited.

The semi-classical description of the fragmentation process, on the
other hand, has been quite successful phenomenologically
\cite{Art74,And79,And83,Art84,And83a,Sjo86,Sjo06,Sjo14,Sch51}.  Following Fig.\ 10 of \cite{Art74}, we show in Fig. 1 an example
of the fragmentation of a $u$-$(ud)$ flux tube with an invariant mass of
$\sqrt{s}$=8.65 GeV as
implemented in
 the Lund model in PYTHIA 6.4 \cite{Sjo06}.
One
envisages in Fig. 1(b) a leading quark  $u$ pulling apart from a leading 
diquark $ud$ at high energies.  The vacuum is so polarized that ordered
$\bar q_{i}$-$q_i$ pairs are produced at vertices   $C_i$  inside the tube via the
Schwinger pair-production mechanism \cite{Sch51}.  The interaction of
a produced quark $q_i$ with an antiquark  $\bar q_{i+1}$ produced in the adjacent vertex
leads to the production of a hadron (most likely a meson) as a yo-yo $h(q_i \bar q_{i+1})$ state and the fragmentation of the flux tube.  The rapidities of the
produced chain of hadrons are ordered along the spatial longitudinal
$z$-axis, and in time.

\begin{figure}[h]

\includegraphics[scale=0.43]{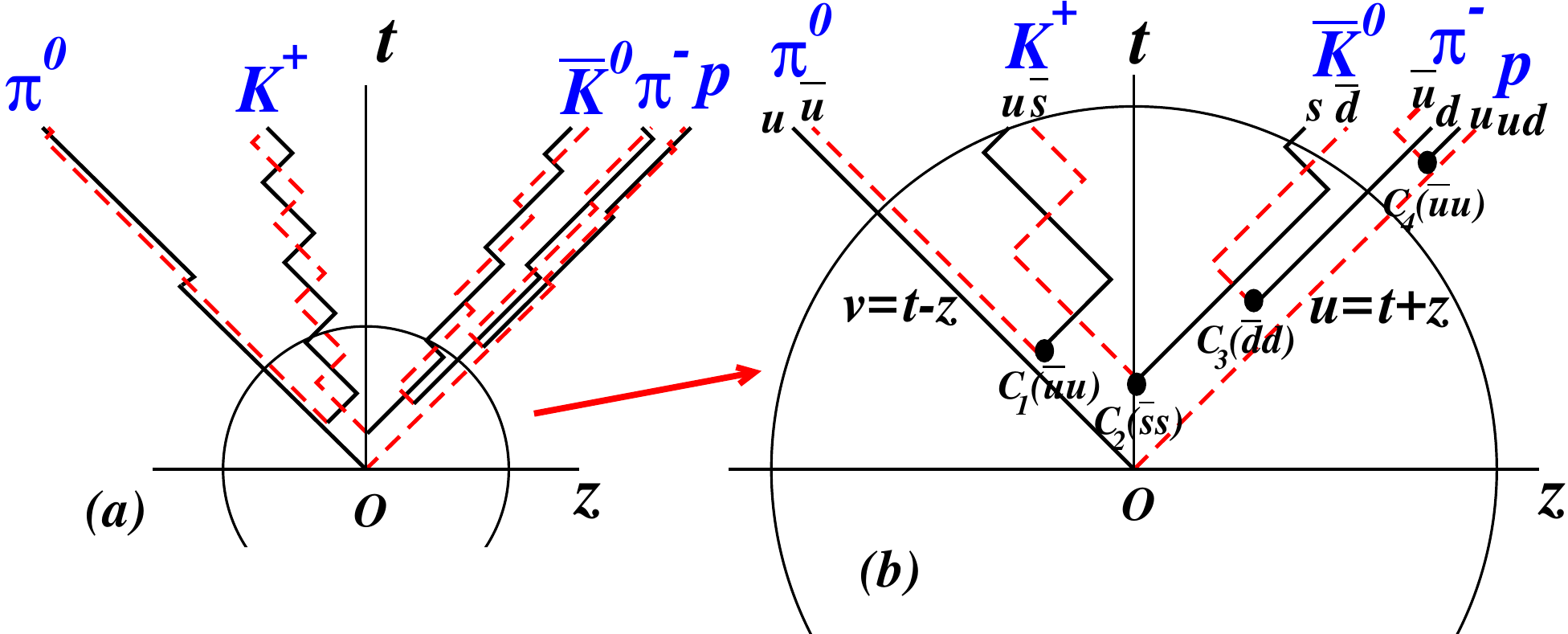}
\caption
{
\label{etaphi} 
An example of the space-time diagram in the
fragmentation of a $u$-$(ud)$ flux tube with an invariant mass of
$\sqrt{s}$=8.65 GeV, obtained with PYTHIA 6.4 \cite{Sjo06}.  Quark
trajectories are shown as solid lines, and antiquark (or diquark) trajectories as
dashed lines. 
     Fig.1(b) is an expanded view of Fig.1(a).
  As the leading $u$ quark  moves in the
  negative $z$-direction and the leading diquark $ud$  in the
  positive $z$-direction with nearly the speed of light, (massless)
  ${\bar q}_i$-$q_i$ pairs will be produced at vertices $C_i(u_i,v_i)$ lying
 approximately on a curve of $v$=$v(u)$, where $u=t+z$ and $v=t-z$.  
 Subsequently, an antiquark ${\bar q}_i$
  produced at one vertex $C_i$ interacts with the quark $q_{i+1}$
  produced in the adjacent vertex $C_{i+1}$ to form a yo-yo $h({\bar q}_i  q_{i+1})$ state
  which can be considered to represent a produced hadron.  
}
\end{figure}

While the semi-classical description of the flux-tube fragmentation
process has been quite successful phenomenologically 
\cite{Art74,And79,And83,Art84,And83a,Sjo86,Sjo06,Sjo14,Sch51}, 
a greater insight and a deeper understanding may be gained if the accompanying 
many-body 
hadronization dynamics  can be
tested on an event-by-event basis.  In this respect, it is
interesting to note that the semi-classical description provides a set
of powerful tools to predict  explicitly the many-body dynamics of the
production process.  Specifically, the process of flux-tube
fragmentation is characterized by (i) a rapidity-space-time ordering
of produced hadrons, (ii) the conservation of charge, flavor, and
momentum for all $q$-$\bar q$ pair productions, and thus (iii) a
complete correlation of all produced hadrons (mostly mesons and
predominantly pions) in an orderly manner.  If the particle production
and hadronization process indeed follows such a semi-classical
description, the above properties may allow the
reconstruction of the configuration of the pair production space-time
vertices of the flux tube.  A chain of produced hadrons at the
instances of fragmentation may be exhibited on an event-by-event
basis, in exclusive (or nearly exclusive) measurements in which the
momenta of almost all hadrons have been measured.  Following
Ref.\ \cite{Won91b}, we suggest the use of event-by-event exclusive
measurements to investigate the space-time many-body dynamics in
flux-tube fragmentation in high energy $e^+e^-$ annihilations, $pp$,
$pA$, and light $AA$ collisions.

Besides testing the basic contents of the semi-classical description
of the many-body correlations, the measurements as proposed here will
serve many other useful purposes.  It is worth noting that  direct experimental evidences for the number of flux tubes and the nature of the flux tubes   in a $pp$ collision are still lacking.
In the conventional Lund Model \cite{Sjo06},  a nucleon-nucleon collision is
generally considered to consist of two colorless   flux
tubes.   
 However, 
 there are other models of particle production in terms of a more  general
color-singlet and  non-color-singlet flux tubes in a $pp$ collision, with different flux tube fragmentation properties.  
 With an increase in the typical momentum transfer (or resolution power)
in nucleon-nucleon (and nucleus-nucleus) collisions 
at very high energies,  the probability of finding gluons in a hadron (or a nucleus)  increases, and the hadron can be described not only by the valence quarks and diquarks as in the Lund Model but also by a color object of predominantly gluons in the form of a color-glass condensate \cite{Mc94,Mc01,Mc10a,Mc10b,Kov95,Gei09}.  The fields in the region between the colliding gluon partons in a high energy $pp$   collision can be described by  color flux tubes in the form of a glassma consisting of a region of non-zero color electric and color magnetic fields \cite{Mc94,Mc01,Mc10a,Mc10b,Kov95,Gei09}.  The decay  of the glassma flux tubes may arise not from  pair production but from the evolution of 
the  classical Yang-Mills fields until the density is so low as to reach the hadronization limit \cite{Mc10a,Mc10b}.  The number of flux tubes in such a decay will depend on the gluon transverse density of the colliding hadrons and is yet to be empirically determined.   The property of the fragmentation process of these tubes will need to be further investigated both theoretically and experimentally.

The explicit
reconstruction of space-time dynamics of the many-body correlations
proposed here may allow the identification and the enumeration of the  fragmenting flux tubes to discrimnate between different models.
If these flux tubes can be successfully identified in $pp$ collisions,
they will provide additional experimental information on the 
nature of the flux tube,  and the
laws that
govern the partition, the interference, and the interaction of the 
flux tubes.  Furthermore, in more complicated $pA$ and $AA$ collisions
involving a proton and light nuclei, the number of participating
nucleons increases, and the dynamics of  the flux
tube fragmentation  will display the effects of multiple
collisions, the effects of the interaction between flux tubes, and the
possibility of the merging of the flux tubes into ropes
\cite{Feo08}.   It will stimulate future  research on
the quantum field description of the particle
production process.

It is important to note that successes in the event-by-event studies
carried out for flux-tube fragmentation will pave the way for future
event-by-event investigations of the dynamics of other
reaction mechanisms such as the hard-scattering process
\cite{Bla74,Ang78,Fey78,Owe78,Rak13,Eli96,Won12}, direct fragmentation process
\cite{Won80}, color-glass condensate \cite{Mc94,Mc01,Mc10a,Mc10b,Kov95,Gei09}, EPOS model
\cite{Wer06}, and Landau hydrodynamics
\cite{Lan53,Won08,Mur04,Ste05,Sen15,Won14}.  Each of these mechanisms
exhibits specific space-time dynamics on an event-by-event basis and
are important in different collision energies and $p_T$ domains.  For
example, if the collision dynamics in a $pp$ collision follow the
space-time description of Landau hydrodynamics, the exact solution of
Landau hydrodynamics in one space and one time directions
\cite{Lan53,Won08,Won14} will predict the occurrence of a single tube
of particles ordered in rapidity, instead of the two flux tubes in the
flux-tube fragmentation model.  If the hydrodynamic bulk matter
freezes out into hadrons statistically and incoherently, then there will
be no azimuthal back-to-back correlations and no flavor correlations
between two hadrons adjacent in rapidity.  In another example, a hard
scattering process of two partons in a $pp$ collision
\cite{Bla74,Ang78,Fey78,Owe78,Rak13,Eli96,Won12} will result in a dijet consisting 
of two azimuthally back-to-back cones of produced hadrons
on an event-by-event basis.  These two cones will appear as 
two  clusters of particles in the $(\eta$, $\phi)$ plane,
centered at $\Delta \phi$$\sim$$\pi$ and 
at 
rapidities (or pseudorapidities) that depend on the  initial 
longitudinal momenta of the colliding partons.  The event-by-event
dynamics of the produced particles can be utilized to study dijet and multiple dijet 
occurrences in $pp$ collisions.
By
identifying  the
particle production mechanism that has occurred in an event, 
one may discriminate different theoretical models and gain 
important pieces of information on the production and
hadronization process.  It is therefore useful to develop methods to
analyze event-by-event dynamics of produced hadrons in order to
extract the wealth of information they provide.

The  importance of the color flux-tube fragmentation process relative to  the hard-scattering process in $pp$
collisions depends on the  collision energy  and the $p_T$ domain
\cite{Won15}.  As the collision energy increases, the hard-scattering
process becomes relatively more and more dominant and the 
transverse momentum boundary $p_{Tb}$ that separates the flux-tube
fragmentation region from the hard-scattering region
recedes to a smaller value of $p_T$ \cite{Won15}.  There is some
evidence of the dominance of flux-tube fragmentation for low-$p_T$
hadron production at central rapidity in $pp$ collisions at energies
$\sqrt{s}$ = 6 to 200 GeV from the study of two-hadron correlations
data of the STAR and the NA61/SHINE Collaborations \cite{Won15,
  STAR06twopar,Por05,Tra11,Ray11,TraKet11,NA61,Mak15,Gaz15,Lar15,Ser15}.
The $pp$ and $AA$ collisions in the NA61/SHINE and NICA energy range
are particularly suitable for event-by-event studies of the space-time
dynamics in flux-tube fragmentation.

It has recently been suggested that a knotted/linked network of flux
tubes may be formed in the early cosmological evolution in a QCD-like
phase transition that may provide the source for the inflation of the
universe \cite{Ber15}.  The knotty network of flux tubes explains why
our universe has exactly three (large) spatial dimensions because flux
tube knots are topologically stable only in exactly three space
dimensions.  In the late stages of the knotty inflation, the flux tube
will fragment into smaller entities and the standard cosmological
evolution begins.  It is therefore of interest to study the space-time
dynamics of the fragmentation of a flux tube to assist possible
identification of flux-tube remnants in knotty inflation cosmological
evolution.

This paper is organized as follows.  In Section II, we discuss the
event-by-event fluctuations and the underlying average features of the
space-time dynamics that may be described semi-classically.  Within
the semi-classical description of flux-tube fragmentation, we derive
the differential equation relating the forward and backward
longitudinal light-cone coordinates $(u,v)$ of the $q$-$\bar q$
pair-production vertices with the rapidity distribution $dN/dy$ of the
observed hadrons.  The equations can be further reduced into a set of
coupled equations for $u$ and $v$ as a function of $y$.  In Section
III, we prove the property of the space-time-rapidity ordering for a
general hadron rapidity distribution $dN/dy$.  In Section IV, we solve
for the space-time coordinates for many different common rapidity
distributions.  In Section V, we infer the space-time
coordinates from the rapidity distribution of $\pi^-$ obtained by the
NA61/SHINE Collaboration.  We review the experimental signatures for
flux-tube fragmentation in Section VI.  In Section VII, we outline the
procedures to carry out event-by-event exclusive measurements to study
flux-tube fragmentation dynamics.  In Section VIII and Appendix A, we
discuss the experimental challenges in event-by-event studies and the
heavier meson fractions as inferred from experimental data.  In Section
IX, we discuss the useful tools that may assist the reconstruction of  
the  flux tubes in a $pp$ collision.
In Section X, we study how the many-hadron correlations may
distinguish the difference between the inside-outside or
outside-inside cascades in the production of particles in flux-tube fragmentation.  In Section
XI, we present our conclusions and discussions.

\section{Space-time dynamics in flux-tube fragmentation}
\label{sec:1}

It is clear from the outset that on an event-by event basis there will
be high degrees of fluctuations and uncertainties in the dynamical
variables of the produced hadrons in a flux-tube fragmentation.  
Nevertheless, each event will
exhibit distinct dynamical characteristics and obey conservation laws
that will presumably remain the same in different events.  We can illustrate here
some of these fluctuations, uncertainties, and conservation laws.

When basic quantum mechanics and the finite size are taken into
account, the pair production mechanism manifests itself as a tunnelling
phenomenon over a potential barrier \cite{Wan88}.  For a static field,
the probability of pair production in a flux-tube fragmentation begins
to become substantial only when the source quark and antiquark are
separated at a distance greater than $d$, as determined by the masses
$2m$ of the produced particle pair and the string tension $\kappa$
 \cite{And83,Wan88},
\begin{eqnarray}
d \ge \frac{2m}{\kappa}.
\label{eq0}
\end{eqnarray}
There is thus an uncertainty of $\Delta z$=$ d$=$2m/\kappa$ in the
spatial separation of the pair-production vertices due to quantum
mechanics, in contrast to the semi-classical model with massless
quarks where the pair of particles are produced at a single point.
Furthermore, the probability of pair production exhibits oscillations
as a function of the spatial locations, of the order of 10\% percent in
the central region, but substantially larger at the edges (Fig. 2 and
3 of \cite{Wan88}).  When the transverse momentum of the produced
fermion is included, the transverse momentum contributes to the
transverse mass $m_T$ which replaces the mass $m$ in Eq.\ (\ref{eq0}).
Thus, the minimum production separation $d$ depends on the transverse
momentum of the produced pair, and there will be additional
fluctuations arising from the transverse momentum degree of freedom.
In a similar manner, there will  also be fluctuations associated with
the flavor and strangeness degrees of freedom, as well as with the
time-dependence of the pair-producing field.

In spite of these fluctuations, variations, and uncertainties, a 
flux-tube fragmentation will possess the characteristic
pattern of a chain of hadrons, linking together as products of the
flux tube.  They are ordered in rapidity and correlated with their
neighbors through the conservation of momentum, charge, and flavor in
the pair production process.  There will be an average space-time
distribution around which the space-time coordinates of the produced
$q$-$\bar q$ pair vertices  will fluctuate.  This average distribution can be identified
as what can be obtained in the semi-classical description.  For this
purpose, we seek a relation between the space-time coordinates of the
pair-production vertices and the rapidity distribution of the produced
hadrons in the semi-classical model.

In the quantum field theory of QED2, Casher $et~al.$ \cite{Cas74}
showed that when a charged particle pulls away from its antiparticle
with nearly the speed of light in the center-of-mass system, the
dipole density of produced charged pairs is a Lorentz-invariant
function and the lines of constant produced dipole density of charged
pairs are hyperbolas with a constant proper time $\tau$.  In a
semi-classical picture of the fragmentation of a color flux tube, it
can be shown that if all pair-production vertices of a fragmenting
string fall on the curve of a constant proper production time
$\tau_{\rm pro}$, the rapidity distribution $dN/dy$ of the produced
hadrons will be a constant given by \cite{Won91b,Won94}
\begin{eqnarray}
\frac{dN}{dy}=\frac{\kappa \tau_{\rm pro}}{m_T},
\label{1}
\end{eqnarray}
where $m_T$=$\sqrt{m^2+p_T^2}$ is
the transverse mass of a produced hadron (mostly likely a pion) with a
rest mass $m$.  Thus the presence of a rapidity plateau is an
indication of the occurrence of the $q$-$\bar q$ production vertices
along a curve of constant proper time, $\tau= \tau_{\rm pro}$,
represented by $uv$=$\tau_{\rm pro}^2$.  Here, the forward and
backward light-cone coordinates $(u,v)$ are related to the space-time
coordinates $(t,z)$ by
\begin{subequations}
\begin{eqnarray}
u&&=t+z,\\
v&&=t-z .
\end{eqnarray}
\end{subequations}
As $(u,v)$ and $(t,z)$ are related by a simple linear transformation,
we shall call both $(u,v)$ and $(t,z)$ as the space-time coordinates.
We can show further that in this case with boost invariance, the
produced particles are ordered in rapidity and space-time
\cite{Won94}.  The rapidity-space-time ordering of produced particles
means that in the center-of-mass system, particles with a greater
magnitude of rapidity $|y|$ are produced at a greater magnitude of the
longitudinal coordinate $|z|$ and at a later time $t$.
 
Experimentally, the observed shape of $dN/dy$ of produced hadrons in
$pp$ collisions is closer to a Gaussian distribution rather than a
flat plateau distribution \cite{Mur04,Ste05,NA61}.  The $q$-$\bar q$
production vertices are not expected to lie on the curve of a constant
proper time.  How do the space-time coordinates of the $q$-$\bar q$
production vertices depends on the hadron rapidity distribution
$dN/dy$?  If the $q$-$\bar q$ production vertices do not lie on the
curve of a constant proper time, would the property of
rapidity-space-time ordering of hadrons be maintained?  Is the
rapidity-space-time ordering of hadrons a general result that does not
require the occurrence of a plateau structure in the hadron rapidity
distribution $dN/dy$?

To answer these questions, we use $u$ as the independent variable and
represent the locus of space-time light-cone coordinates $(u,v)$ of
the pair-production vertices $C_i(u_i,v_i)$ in Fig. 1 by $v(u)$ as a
function of $u$.  We would like to write down the differential
equation governing $v(u)$ in terms of $dN/dy$.

The rapidity of a hadron yo-yo state $y_i$ is related to the
space-time coordinates of its constituent quark and antiquark that
have been produced at vertices $C_{i-1}$=$(u_{i-1},v_{i-1})$ and
$C_{i}$= $(u_i,v_i)$ by
\begin{eqnarray}
y_i=\frac{1}{2} \ln \frac{ \kappa (u_{i+1}-u_i)}{\kappa (v_i-v_{i+1})}.
\label{2}
\end{eqnarray}
Therefore, we have
\begin{eqnarray}
\frac{\Delta N}{\Delta y}\! =\! \frac{1}{y_{i}-y_{i-1}}\!=\! 2\!\biggr /\!\! \left [ \ln \!\left ( 
\frac{  u_{i}-u_{i+1}}{v_{i+1}-v_i}\right )\! -\!\ln \!\left (\frac{  u_{i-1}-u_i}{v_i-v_{i-1}} \right )\! \right ] .
\nonumber\\
\label{3}
\end{eqnarray}
Upon taking the continuum limits of the above two equations (\ref{2})
and (\ref{3}), we get  \cite{Won94}
\begin{eqnarray}
y&&=-\frac{1}{2}\ln \left ( - \frac{d v}{du}\right ) ,  
\label{4}
\\
 \frac{dN}{dy}&& =\frac{2\kappa}{m_T} \left ( - \frac{dv}{du}\right )^{3/2} \left ( 
\frac{d^2 v}{du^2}\right )^{-1}, 
\label{5b}
\end{eqnarray}
where we have used the relation between $\Delta u_i$=$u_{i+1}-u_{i}$,
$\Delta v_i$=$v_{i+1}-v_{i}$, and the transverse mass $m_{Ti}$ of the
produced hadron, $\Delta u_i (- \Delta v_i)$=$(\Delta u_i)^2 (-
dv/du)={m_{Ti}^2}/{\kappa^2}.$ From Eq.\ (\ref{5b}), the differential
equation that governs $u(v)$ is related directly to the hadron
rapidity $dN/dy$ by
\begin{eqnarray}
{ \frac{d^2 v }{du^2}}
 = \frac{2\kappa}{m_T (dN/dy)} {\left ( - \frac{dv}{du}\right )^{3/2}}.
\label{6}
\end{eqnarray}
If $dN/dy$ is a known function of $y$, then Eqs.\ (\ref{4}) and
(\ref{6}) can be used to solve for $v(u)$.

We can re-write Eqs. (\ref{4}) and (\ref{6}) in a simpler form to
facilitate the solution of $u$ and $v$ for a general $dN/dy$.  From
Eq. (\ref{4}), we have
\begin{eqnarray}
 -  \frac{dv}{du}=e^{-2y}.
\label{7}
\end{eqnarray}
Substituting this into Eq.\ (\ref{6}), we obtain
\begin{eqnarray}
\frac{dy}{du}=\frac{\kappa}{m_T (dN/dy)} e^{-y},
\end{eqnarray}
which gives 
\begin{eqnarray}
\frac{du}{dy}=\frac{m_T } {\kappa}\frac{dN}{dy}e^y.
\label{8}
\end{eqnarray}
Using $dv/dy=(dv/du)(du/dy)$ and Eq.\ (\ref{7}) we obtain the equation
of $dv/dy$
\begin{eqnarray}
\frac{dv}{dy}=-\frac{m_T } {\kappa}\frac{dN}{dy}e^{-y}.
\label{9}
\end{eqnarray}
The set of Eqs.\ (\ref{8}) and (\ref{9}) allow one to solve for the
space-time coordinates $(u,v)$ and $(t,z)$ of the $q$-$\bar q$
pair-production vertices as a function of the rapidity variable $y$,
for a general hadron rapidity distribution $dN/dy$.

\section{Rapidity-Space-Time Ordering of the Produced Hadrons}

We can infer some important gross properties of the
rapidity-space-time ordering from Eqs.\ (\ref{8}) and (\ref{9}).  We
can evaluate $dy/dz$ and we get
\begin{eqnarray}
\frac{dy}{dz} &&= \frac{d y}{d u}\frac{du}{dz} +  \frac{d y}{d v}\frac{dv}{dz}
\nonumber\\
&&= \frac{m_T } {\kappa}\frac{dN}{dy} 2\cosh y.
\end{eqnarray}
Because all factors on the right-hand side of the above equation are
positive, we have
\begin{eqnarray}
\frac{d y}{d z} >0.
\label{14}
\end{eqnarray}
This  says that the rapidities of the produced hadrons are ordered in
longitudinal space.  Those hadrons with a greater rapidity $y$ are
produced at a greater value of the $z$ coordinate.  We can similarly
evaluate $dy/dt$,
\begin{eqnarray}
\frac{dy}{dt} &&= \frac{dy}{du}\frac{dy}{dt}  +\frac{dy}{dv}\frac{dv}{dt}
\nonumber\\
&&= \frac{m_T } {\kappa}\frac{dN}{dy} 2\sinh y,
\end{eqnarray}
which says that 
\begin{eqnarray}
\begin{cases}
dy/d t <0  & {\rm for} ~y<0,\cr
dy/d t = 0 & {\rm for} ~y=0,\cr
dy/d t >0  & {\rm for} ~ y>0.
\label{16}
\end{cases}
\end{eqnarray}
Combining the above results on $dy/dz$ and $dy/dt$ in Eqs. (\ref{14})
and (\ref{16}), we obtain the rapidity-space-time ordering of produced
hadrons.  The hadrons with a greater magnitude of rapidity $|y|$ are
produced at a greater magnitude of the longitudinal coordinate $|z|$
and at a later time $t$.
 
We have thus proved the result of the rapidity-space-time ordering for
a general rapidity distribution $dN/dy$ in a semi-classical
description of flux-tube fragmentation, provided the vertices of the
produced $q$-$\bar q$ pairs lie on a curve of $v$=$v(u)$.  It should
be pointed out that the production of a $q$-$\bar q$ pair is a
stochastic process, whose occurrence has a probability distribution
and fluctuations.  The coordinates of the vertices should fluctuate
about an average distribution on an event-by-event basis.

\section{Solution for the space-time coordinates of 
pair-production vertices for a given $dN/dy$ }

Given a hadron rapidity distribution $dN/dy$ that is a function of
$y$, Eqs.\ (\ref{8}) and (\ref{9}) can be integrated out to give a
parametric representation of the space-time coordinates $(u,v)$ and
$(z,t)$ of the $q$-$\bar q$ production vertices.  They assume
analytical forms for many common hadron rapidity distributions
$dN/dy$, including a boost-invariant, a plateau distribution, and a
Gaussian distribution.

\subsection { Space-time vertex coordinates for a boost-invariant  $dN/dy$}

We consider first the case of a boost-invariant $dN/dy$ that is
independent of the rapidity.  The distribution can be represented by
\begin{eqnarray}
\frac{dN}{dy}=(dN/dy)_0,
\end{eqnarray}
where $(dN/dy)_0$ is $dN/dy$ at $y=0$.
In terms of the average transverse mass $m_T$ and the string constant
$\kappa$, we can introduce $\tau_0$ given as in Eq.\ (\ref{1}) by
\begin{eqnarray}
\tau_0=\tau_{\rm pro}=  \frac{m_T} {\kappa}  (dN/dy)_0.
\label{16}
\end{eqnarray}•
Eqs.\ (\ref{8}) and (\ref{9}) then yield the solution of the 
space-time coordinates
\begin{subequations}
\begin{eqnarray}
u(y) &&=\tau_0 e^y \\
v(y) && =\tau_0 e^{-y},
\end{eqnarray}
\end{subequations}
and 
\begin{subequations}
\begin{eqnarray}
t &&=\tau_0 \cosh y \\
z && =\tau_0 \sinh y.
\end{eqnarray}
\end{subequations}
As a consequence,
\begin{eqnarray}
u(y)v(y)&&=\tau^2 =t^2-z^2=\tau_0^2.
\end{eqnarray}
This is the well-known result that the boost-invariant pair-production
vertices lie, on the average, on a curve of constant proper time,
$\tau=\tau_0=\tau_{\rm pro}$.  It occurs only for the case in the
fragmentation of a flux tube with infinitely large energies.

\subsection{Space-time vertex coordinates for a  plateau rapidity 
distribution $dN/dy$}

For the fragmentation of a flux tube with a finite energy, we consider
next the more realistic case of a rapidity distribution with a plateau
structure within a limited region of $|y|\le y_{\rm max}$,
\begin{eqnarray}
\frac{dN}{dy}=(dN/dy)_0 \Theta(y_{\rm max}-|y|) .
\end{eqnarray}
We can again represent $(m_T/\kappa)(dN/dy)_0$ by $\tau_0$ as in
Eq.\ (\ref{16}).  Equations (\ref{8}) and (\ref{9}) yield the solution \break for
$|y|\ge y_{\rm max}$,
\begin{subequations}
\begin{eqnarray}
u(y) &&=\tau_0 e^y \\
v(y) && =\tau_0 e^{-y},
\end{eqnarray}
\end{subequations}
and 
\begin{subequations}
\begin{eqnarray}
t &&=\tau_0 \cosh y \\
z && =\tau_0 \sinh y.
\end{eqnarray}
\end{subequations}
In this case with a rapidity plateau distribution, the ranges of $t$
and $z$ are limited, with $\tau_0 \le t \le \tau_0 \cosh y_{\rm max}$
and $|z| \le \tau_0 \sinh y_{\rm max}$.
 
\subsection{Space-time vertex coordinates for a  Gaussian $dN/dy$}

We consider the fragmentation of a flux tube which gives rise to the
production of hadrons with the Gaussian rapidity distribution centering
at the rapidity $y_0$,
\begin{eqnarray}
\frac{dN_{\rm FT}}{dy} =N_{\rm FT} \frac{\exp \{-\frac{(y-y_0)^2}{2\sigma^2} \}} {\sqrt{2\pi}\sigma} ,
\label{26}
\end{eqnarray}
where $N_{\rm FT}$ is the integrated number of hadrons (predominantly
pions) produced per flux tube.  The equation for $du/dy$,
Eq.\ ({\ref{8}), becomes
\begin{eqnarray}
\frac{du}{dy}&&=\frac{m_T } {\kappa} \frac{N_{\rm FT}}{\sqrt{2\pi}\sigma} e^{-\frac{1}{2\sigma^2}(y-y_0)^2}e^y.
\end{eqnarray}
This differential equation can be  solved to give the solution
\begin{eqnarray}
u(y) = e^{y_0}\left [\frac{m_T } {\kappa}  \frac{N_{\rm FT}e^{\sigma^2/2}}{2} \left \{ 1 + {\rm erf}(\frac{y-y_0-\sigma^2}{\sqrt{2}\sigma}) \right \}\right ]\! .~~~~~~
\end{eqnarray}
The equation for $dv/dy$, Eq.\ (\ref{9}),  is
\begin{eqnarray}
\frac{dv}{dy}&&=-\frac{m_T } {\kappa} \frac{N_{\rm FT}}{\sqrt{2\pi}\sigma} e^{-\frac{1}{2\sigma^2}(y-y_0)^2}e^{-y},
\end{eqnarray}
which can also be solved  to give the solution
\begin{eqnarray}
v(y) = e^{-y_0}\left [\!\frac{m_T } {\kappa} \frac
{N_{\rm FT} e^{\sigma^2/2} } {2}  \left \{ 1\! -\! {\rm erf}(\frac{y-y_0+\sigma^2}{\sqrt{2}\sigma}) \right \}\!\right ]\!. ~~~~
\end{eqnarray}
We can form $t=(u+v)/2$ and $z=(u-v)/2$ to yield
\begin{eqnarray}
t=  \frac{m_T } {\kappa}
 \frac{N_{\rm FT} e^{\sigma^2/2} }{4} 
 && \biggl  [ e^{y_0} \left \{ 1 + {\rm erf}(\frac{y-y_0-\sigma^2}{\sqrt{2}\sigma}) \right \}
\nonumber\\
+&& e^{-y_0} \left \{ 1 -  {\rm erf}(\frac{y-y_0+\sigma^2}{\sqrt{2}\sigma}) \right \}
\biggr ],
\label{31}
\end{eqnarray}
\begin{eqnarray}
z=  \frac{m_T } {\kappa}
 \frac{N_{\rm FT} e^{\sigma^2/2} }{4} 
 && \biggl [e^{y_0} \left \{ 1 + {\rm erf}(\frac{y-y_0-\sigma^2}{\sqrt{2}\sigma}) \right \}
\nonumber\\
-&& e^{-y_0} \left \{ 1 -  {\rm erf}(\frac{y-y_0+\sigma^2}{\sqrt{2}\sigma}) \right \}
\biggr ].
\label{32}
\end{eqnarray}
The above two equations give the parametric representation of the
space-time coordinates $t$ and $z$ of the produced $q$-$\bar
q$ pairs leading to the production of hadrons with the Gaussian
rapidity distribution Eq.\ (\ref{26}).

\section{Space-time coordinates of $q$-$\bar q$ production vertices 
inferred from NA61/SHINE $dN_{pp}(\pi^-)/dy$ data}

The results in the last section indicates that the space-time
coordinate of $q$-$\bar q$ production vertices are directly related to
the hadron rapidity distribution.  We need the experimental hadron
rapidity distribution data $dN/dy$ to infer the space-time coordinates
of $q$-$\bar q$ production vertices in flux-tube fragmentation.  As
the experimental $dN/dy$ data represent the rapidity distribution
averaged over many events, the space-time coordinates inferred from
the experimental $dN/dy$ data should also be considered as 
space-time coordinates averaged over an ensemble of events.

In $pp$ collisions at energies investigated by the 
NA61/SHINE Collaboration,  the flux-tube fragmentation process occurs in the
central rapidity region, whereas target and projection fragmentation
processes take place near the projectile and target rapidities,
respectively.  As a consequence, positive (positively charged) hadrons
can be produced in all three regions of the
rapidity space in $pp$ collisions.  On the other hand, low-$p_T$ negative (negatively
charged) pions are unlikely to be produced in the projectile and
target fragmentation regions in $pp$ collisions, and can be
appropriately considered to originate predominantly from the flux-tube
fragmentation process in the central rapidity region without contributions from projectile and target fragmentations.

The NA61/SHINE Collaboration has measured the transverse and
longitudinal spectra of negative pions in the collision of a proton on
a fixed proton target at $p_{\rm lab}=$ 20, 31, 40, 80, and 158 GeV/c
\cite{NA61}.  The transverse spectra of low-$p_T$ pions can be
represented by $dN/m_T d m_T\propto \exp\{ - m_T/T\}$ where $T$ ranges
from 150 to 160 MeV \cite{NA61}.  After integrating over the
transverse spectra, the NA61/SHINE $\pi^-$ data of the rapidity
distribution for different $p_{\rm lab}$ are shown in
Fig.\ \ref{dndynn}.

Our event-by-event study has been motivated to obtain a space-time description of the $pp$ collisions process
 for which direct experimental information remains lacking.
Different models have been presented for such a description with varying degrees of successes, but they will need to be tested with direct event-by-event measurements which we envisage here.

The most common description for the flux tube fragmentation in a $pp$ collision is given in terms by the Lund Model \cite{Art74,And79,And83,Art84,And83a,Sjo86,Sjo06,Sjo14}.
It is assumed that in  the relatively lower energy region in $pp$ collisions,
the valence quarks and  valence diquarks of the colliding protons 
dominates the soft particle production process.
By color conservation, the only simpest colorless objects formed by the valence consituents of one nucleon and the valence consituents of the other nucleon after the collision are two 
colorless flux tubes formed by    
 the  valence quark of one proton and the valence diquark of the other colliding proton.    Each colorless flux tube undergoes fragmention in the same way as  
  in an $e^+$-$e^-$ annihilation.
At higher $pp$ (and $AA$) collision energies, there are other models of particle production in terms of a more  general
color-singlet and  non-color-singlet flux tubes involving also gluons in additon to  valence quark and valence diquarks, with different flux tube fragmentation properties.  
For example,
 at high-multiplicity
events at very high energies, the possibility of forming more than two
flux tubes has been suggested \cite{Sjo86}.
In another  example  at very high energies,  the probability of finding gluons in a hadron (or a nucleus)  increases, and the hadron can be described  by a color object of predominantly gluons in the form of a color-glass condensate in additon to valence quark and valence diquarks \cite{Mc94,Mc01,Mc10a,Mc10b,Kov95,Gei09}.  The fields in the region between the colliding gluon partons after a high energy $pp$   collision can be described by  glassma flux tubes  \cite{Mc94,Mc01,Mc10a,Mc10b,Kov95,Gei09} whose number depends on the gluon transverse density of the colliding hadrons.   The longitudinal expansion  of the glassma flux tubes as the classical Yang-Mill field evolves to lower densities  leads to the fragmentation of the tubes. 

\begin{figure}[h]
\hspace*{0.0cm}
\includegraphics[scale=0.45]{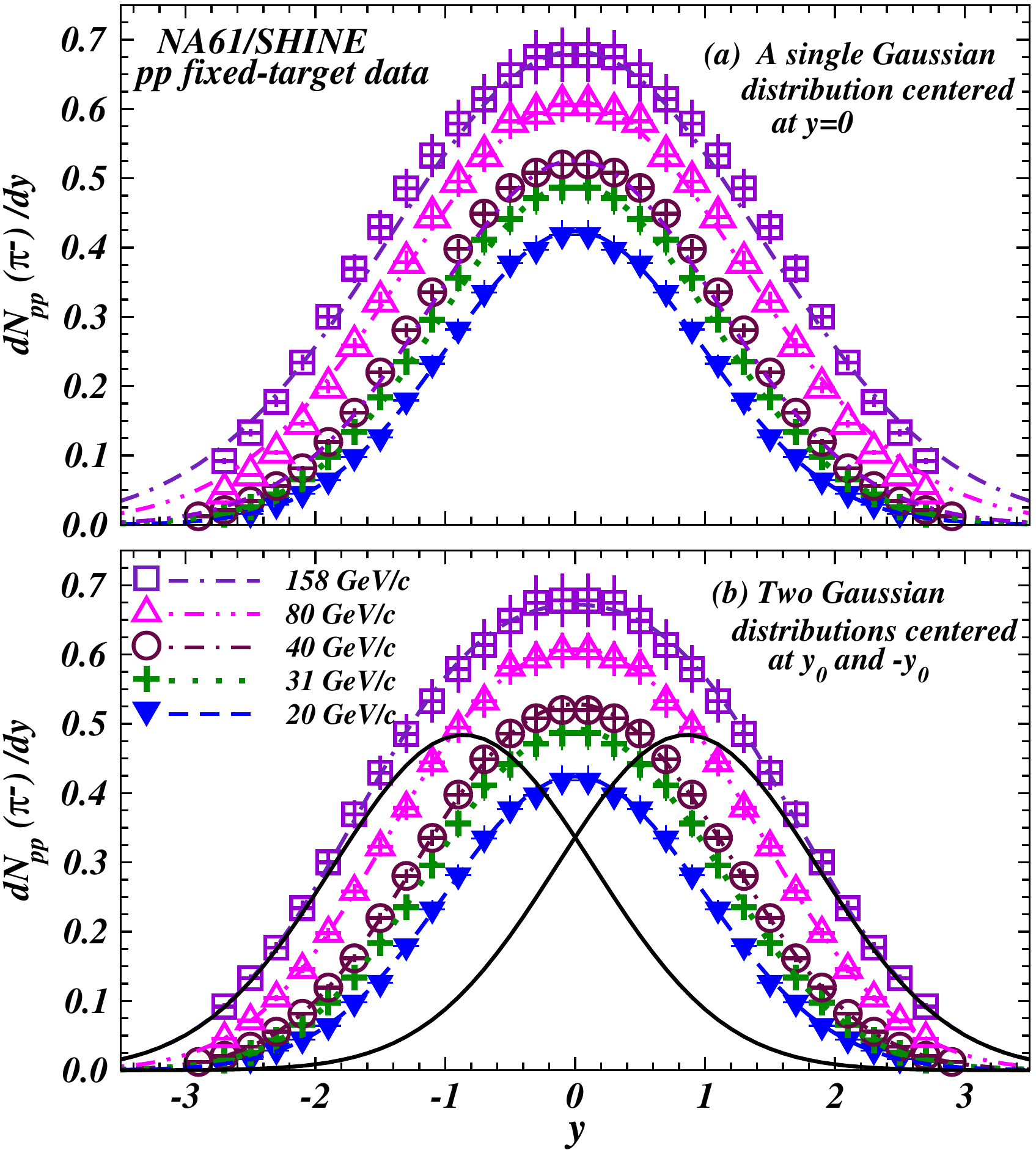}
\caption{ (Color online).  The data points are NA61/SHINE
  $dN_{pp}(\pi^-)/dy$ data for different $p_{\rm lab}$ \cite{NA61}.
  The curves in Fig.\ (a) give the fits to the data points using a
  single Gaussian rapidity distribution, Eq.\ (\ref{eq32}), centering
  at $y=0$, with parameters in Table I.  The curves in Fig.\ (b)
  give fits to the data points using the sum of two displaced Gaussian
  distributions, Eq.\ (\ref{34}), centering at $+y_0$ and $-y_0$, with
  parameters in Table II.  The two solid curves in Fig.\ (2b) give the
  two separate Gaussian distributions at $+y_0$ and $-y_0$ for the
  case of $p_{\rm lab}$=158 GeV.  }
\label{dndynn}
\end{figure}

As the number of flux tubes and their properties in 
a $pp$ collision are not yet known experimentally,   
we shall analyze the production of the low-$p_T$ negative pions  by  assuming two possible scenarios with the occurrence of
(I) a single flux tube, or (II)  two differently-displaced  flux tubes.   The scenarios of many more flux tubes can be similarly generalized.
The proper description to distinguish between the different scenarios may need to rely on 
experimental information that will  be forthcoming
in  the event-by-event analysis.

With the assumption of the  scenario (I) of a single flux tube, we can represent the rapidity distribution 
by a single Gaussian distribution
\begin{eqnarray}
\frac{dN_{pp}(\pi^-)}{dy}=\frac{\langle \pi^- \rangle }{\sqrt{2\pi}\sigma} \exp\{-\frac{y^2}{2\sigma^2}\}.
\label{eq32}
\end{eqnarray}
The fits of  the experimental
NA61/SHINE $\pi^-$ rapidity distribution data 
in terms of a single-Gaussian distribution
are shown in Fig.\ 2(a).
The fitting parameter  $\langle \pi^- \rangle$ is the average number of  $\pi^-$ produced per $pp$ collision,
related to the total number of pions produced per flux tube $N_{\rm FT}$  by 
\begin{eqnarray}
N_{FT}\sim 3\langle \pi^-\rangle /2,
\label{33}
\end{eqnarray}
 and $\sigma$ is the standard deviation of the distribution.
They are listed 
 as a function of the incident $p_{\rm lab}$  in Table I.

\begin{table}[h]
\caption { Parameters of the single-Gaussian fit (Eq. (\ref{eq32})) 
to the NA61/SHINE $\pi^-$ rapidity data in $pp$ collisions \cite{NA61}. } 
\vspace*{0.2cm}
\begin{tabular}{|c|c|c|c|}
\cline{1-4}
     $p_{\rm lab}$ (GeV/c)    &  $\sqrt{s}$ (GeV)  &  $\langle \pi^-\rangle $  &  $\sigma$  
  \\
\cline{1-4}
20   &   6.25   & 1.04$\pm$0.05   & 0.98$\pm$0.030  
\\
31   &   7.75  & 1.28$\pm$0.06  & 1.04$\pm$0.035
\\
40  &  8.77  & 1.45$\pm$0.05   & 1.10$\pm$0.044 
\\
80  &  12.33  & 1.99$\pm$0.08   & 1.30$\pm$0.052 
\\
158  &  17.27  & 2.45$\pm$0.13   & 1.43$\pm$0.065  
\\ \hline
\end{tabular}
\end{table}

To go from the rapidity distribution to the average space-time coordinates of the pair-production vertices, we need a model to relate these quantities. 
For definiteness, we shall provide predictions on the space-time behavior based on  the semi-classical description of the particle production 
in a flux tube fragmentation in the Lund Model  \cite{Art74,And79,And83,Art84,And83a,Sjo86,Sjo06} in Section IVC, so that experimental deviations from such a semi-classical description, if any,  will provide impetus for a search for the other alternative descriptions and models.  
We assume a string tension $\kappa=1$ GeV/fm, and the average
transverse mass $m_T=2T$ from the slope parameters $T$ of the
transverse spectra in \cite{NA61,Mak15,Gaz15,Lar15,Ser15}. 
From Eqs.\ (\ref{26})-(\ref{32}), 
 we obtain the space-time
coordinates of $q$-$\bar q$ production vertices for different $p_{\rm
  lab}$ shown in Fig.\ \ref{uv}(a), under the assumption of scenario (I) of a single flux tube
 in $pp$ collisions.  As $p_{\rm
  lab}$ increases, the curve of the space-time coordinates rises
higher in $t$ and extends farther in $|z|$

\begin{figure}[h]
\includegraphics[scale=0.43]{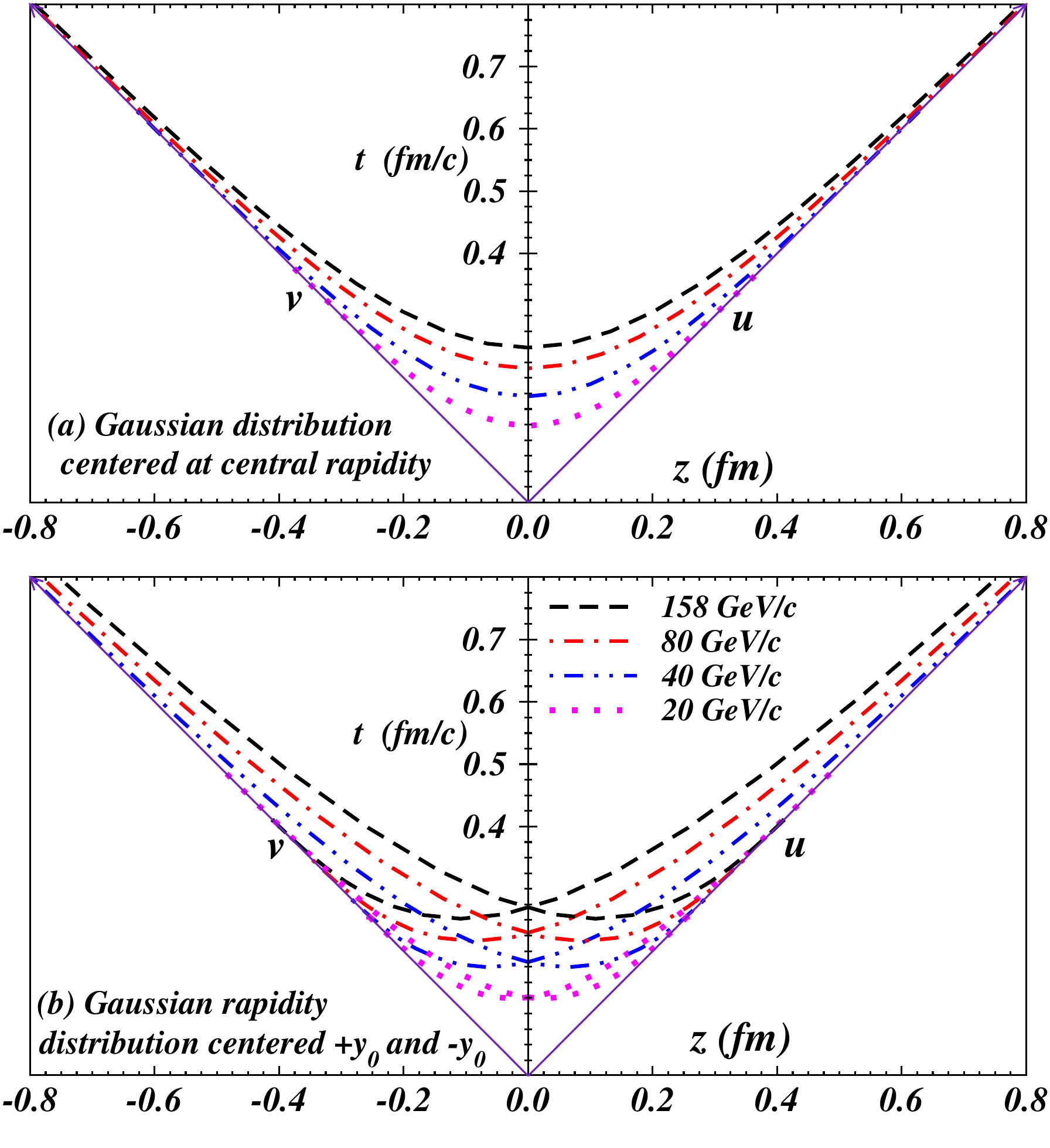}
\caption{ (Color online). The locus of average space-time coordinates
  of $q$-$\bar q$ production vertices for different $p_{\rm lab}$ as
  inferred from the NA61/SHINE $\pi^-$ rapidity distribution data.
  (a) under the assumption (I) of the occurrence of a single flux tube and a 
  Gaussian rapidity distribution, Eq.\ (\ref{eq32}), centering at
  zero, and (b) under the assumption (II) of two flux tubes and two displaced Gaussian
  rapidity distributions, Eq.\ (\ref{34}), centering at $+y_0$ and
  $-y_0$.  }
\label{uv}
\end{figure}

Previously, with an additional parameter,
  the NA61/SHINE Collaboration partitions
the $\pi^-$ rapidity distribution as the sum of two symmetrically displaced
Gaussian distributions centering at $y_0$ and $-y_0$ \cite{NA61},
\begin{eqnarray}
  \frac{d N_{pp}(\pi^-)}{d y} = 
  \frac{\langle \pi^- \rangle  }{2}&&\biggl \{  \frac{1}{\sqrt{2\pi}\sigma} 
  \exp\left(-\frac{(y-y_0)^2}{2\sigma^2}\right)
\nonumber\\
&&
  +\frac{1}{\sqrt{2\pi}\sigma} \exp\left(-\frac{(y+y_0)^2}{2\sigma^2}\right)
  \biggr \} .
\label{34}
\end{eqnarray}
The values of the average number of $\pi^-$ produced  per $pp$ collision  $\langle
\pi^-\rangle$, the rapidity displacement $y_0$, and the standard deviation $\sigma$ of the rapidity distribution    are tabulated for different values of $p_{\rm lab}$
in Table
II (from Table 5 of Ref.\ \cite{NA61}).    The fits to the rapidity
distributions using two displaced Gaussian distributions in
Eq.\ (\ref{34}) are shown in Fig.\ 2(b).  The rapidity distributions
with two displaced Gaussian distributions give a better agreement than
a single Gaussian at the tail regions.

Comparing Table I and II, one notices that $\langle \pi^- \rangle$ are
nearly the same in the two tables.  For the case of two displaced
Gaussian distributions, the displacement $y_0$ increases substantially
as $p_{\rm lab}$ increases, but the standard deviation $\sigma$
changes only slightly.  The increase in $y_0$ for two displaced
Gaussian distributions shows up as an increase in $\sigma$ for a
single Gaussian distribution.

\begin{table}[h]
\caption { Parameters of the two-displaced-Gaussian fit (Eq. (\ref{34})) 
to the NA61/SHINE $\pi^-$ rapidity data in $pp$ collisions, from Ref.\ \cite{NA61}. } 
\vspace*{0.2cm}
\begin{tabular}{|c|c|c|c|}
\cline{1-4}
     $p_{\rm lab}$ (GeV/c)    &  $\langle \pi^-\rangle $  &  $\sigma$   & $y_0$
  \\
\cline{1-4}
20   &   $1.047 \pm 0.051$ & $0.921 \pm 0.118$ & $0.337 \pm 
0.406$ 
\\
31   &  $1.312 \pm 0.069$  & $0.875 \pm 0.050$ & $0.545 \pm 
0.055$ 
\\
40  &  $1.478 \pm 0.051$ &  $0.882 \pm 0.045$ & $0.604 \pm 
0.044$
\\
80  & $1.938 \pm 0.080$  & $0.937 \pm 0.019$ & $0.733 \pm 
0.010$ 
\\
158  &  $2.444 \pm 0.130$  & $1.007 \pm 0.051$ & $0.860 \pm 
0.021$
\\ \hline
\end{tabular}
\end{table}

The success in the description with  two displaced Gaussian rapidity distributions suggests the alternative scenario (II) 
of two flux tubes 
 in a $pp$ collision,
in which 
the fragmentation process takes place between the diquark of one
nucleon and a valance quark of the other nucleon.
The displaced Gaussian  rapidity distribution 
arises from  
the asymmetry between the diquark and the
valence quark in  each flux tube. 
For this case, 
Eqs.\ (\ref{31}) and (\ref{32}) give the parametric representation of
the average space-time coordinates $(t,z)$ of the vertices of
$q$-$\bar q$ pair production in
Fig.\ \ref{uv}(b).  As one observes, the space-time coordinates of the pair-production vertices
in  
one of the two flux tubes in a $pp$ collision is asymmetrical with respect
to the $t$ axis, arising from the rapidity  displacement $y_0$.  As the collision momentum $p_{\rm lab}$ increases, the magnitude  of
the  rapidity  displacement $y_0$ increases, and the asymmetry increases.  The space-time
coordinates of the pair-production vertices 
 rises up as a function of $t$
and are shifted to greater values of $z$,
There is another symmetrically displaced flux tube with hadron
rapidity displacement at $-y_0$.  The corresponding space-time
coordinates of the pair-production vertices can be obtained by a
reflection with respect to the $t$ axis.

It should be emphasized again that what we have shown in
Fig.\ \ref{uv}(a) or \ref{uv}(b) represents the behavior of the
space-time coordinates averaged over an ensemble of events.  The
production of a $q$-$\bar q$ pair is a stochastic process with
fluctuations.  The event-by-event locations of space-time coordinates of $q$-$\bar q$
production vertices  are expected to
fluctuate around the average space-time coordinates obtained in Fig.\ 3.

\section{Experimental Two-hadron Correlation Signature of Flux-Tube Fragmentation}

In a flux-tube fragmentation, the production of quark-antiquark pairs
along a color flux tube precedes the fragmentation of the tube.
Because of local conservation laws, the production of a $q$-$\bar q$
pair will lead to correlations of adjacently produced hadrons (mostly
pions).  On account of the rapidity-space-time ordering of produced
mesons in a flux-tube fragmentation discussed in Section III,
adjacently produced mesons can be signaled by their rapidity
difference $\Delta y$ falling within the window of $|\Delta y | $$<$$
1/(dN/dy)$.  Therefore, the local conservation laws of momentum,
charge, and flavor will lead to a suppression of the angular
correlation function $dN/(d\Delta \phi\, d\Delta y)$ for two hadrons
with opposite charges or strangeness on the near side at $(\Delta
\phi, \Delta y) \sim$ 0, but an enhanced correlation on the
back-to-back, away side at $\Delta \phi \sim \pi$, within the window
of $|\Delta y |$$ <$$ 1/(dN/dy)$.  These properties can be used as
signatures for the fragmentation of a color flux tube \cite{Won15}.

For $pp$ collisions at $\sqrt{s_{pp}}$=200 GeV, the STAR Collaboration
found that if one separates the transverse momentum regions by the
boundary $p_{Tb}$=0.5 GeV/c, the two-hadron angular correlation
pattern, $\Delta \rho/\sqrt{\rho_{\rm ref}} \propto dN/d\Delta \phi
d\Delta \eta$, for two oppositely charged hadrons in the domain below
$p_T$$<$$p_{Tb}$ is distinctly different from the pattern in the
domain above $p_T$$>$$p_{Tb}$
\cite{STAR06twopar,Por05,Tra11,Ray11,TraKet11}.  One finds a
suppression at $(\Delta\eta, \Delta \phi)$$\sim$0 but an enhancement at
$(\Delta\eta$$\sim$0,$\Delta \phi$$\sim$$ \pi)$.  Such a pattern is
consistent with the theoretical correlation pattern for two oppositely
charged hadrons for a flux-tube fragmentation \cite{Won15}.  In the
higher-$p_T$ domain of $p_{Tb}>$0.5 GeV/c, the correlation pattern
is consistent with  the hard-scattering process with the production of two jets
(minijets) \cite{Won15}.

The NA61/SHINE Collaboration has reported the angular correlation for
two hadrons with opposite charges and $p_T$$<$1.5 GeV/c, for $pp$
collisions at $\sqrt{s_{pp}}$ = 6$-$17 GeV
\cite{Mak15,Gaz15,Lar15,Ser15}.  The experimental correlation patterns
show a suppression at $(\Delta \phi, \Delta \eta)\sim$ 0, and an
enhancement at $(\Delta\eta$$\sim$0,$\Delta \phi$$\sim$$ \pi)$,
indicating the dominance of the flux-tube fragmentation in the
low-$p_T$ domain.  Such a two-hadron correlation pattern remains
unchanged up to $p_T$=1.5 GeV/c \cite{Mak15}. The pattern of the
NA61/SHINE two-hadron correlation was also shown to be qualitatively
consistent with the EPOS model \cite{Wer06,Mak15}.  Although there are
many different processes and diagrams in the EPOS model, it is likely
that the dominant process responsible for such a two-hadron
correlation pattern is the flux-tube fragmentation part of the EPOS
model.

\section{ Event-By-Event Exclusive Measurements to Study 
Flux-Tube Fragmentation Dynamics}

The comparison of the theoretical signatures with the experimental
two-hadron correlation data in the last section reveals that there are
$p_T$ regions in $pp$ collisions in which the flux-tube fragmentation
process may dominate.  We are therefore encouraged to examine the
event-by-event dynamics of flux-tube fragmentation in these kinematic
regions, in order to extract the wealth of information in the
flux-tube fragmentation process \cite{Won91b}. 
If the semi-classical description \cite{Art74,And79,And83,Art84,And83a,Sjo86,Sjo14} is a reasonable concept, then
 the local conservation laws and the
rapidity-space-time ordering of hadrons provide
a set of powerful tools  to exhibit  the
chain of produced hadrons and to reconstruct the
space-time coordinates of the $q$-$\bar q$ production vertices.

For an $e^+e^-$ annihilation for which a single flux tube is produced,
one can perform sphericity and thrust analysis to locate the
longitudinal axis with respect to which the transverse masses and the
rapidities of detected hadrons are determined.  
In a $pp$ collision, the conventional   description of flux tube fragmentation is given in terms 
of  
two flux tubes, formed by the quark of one nucleon with the diquark of
the other nucleon and vice versa \cite{Sjo06}.  In another model of flux tubes 
in $pp$ collisions, the flux tubes arises from gluons in the color-glass condensate, and the number
of flux tubes depend on the gluon transverse densities of the colliding hadrons  \cite{Mc94,Mc01,Mc10a,Mc10b,Kov95,Gei09}.

For definiteness of our analysis
in $pp$ collisions,  we shall again follow the Lund semi-classical description in terms of the fragmentation of two colorless  flux tubes so that deviations from such a semi-classical description  will provide impetus for a search for alternative descriptions.  
The beam axis can be selected as the longitudinal axis with respect to
which the transverse masses and rapidities are determined.
The centers of momentum of the two
flux tubes need not be the same.  Thus, it will be necessary to
order the two sets of hadrons according to their rapidities and
decompose them as belonging to one set or the other of the fragmenting
flux tubes.  Each set of hadrons needs to be checked for energy,
charge, and flavor conservation.  The situation is not unlike the
solution of a jig-saw puzzle where one makes use of the conservation
of transverse momentum, flavor, and charge as hints to match the
adjacent hadrons so as to link each flux tube by itself.  By such a
separation, one has two sets of hadrons each of which is likely to
arise from the fragmentation of a single flux tube.  We shall explore
the tools that may facilitate the separation of two flux tubes in
Section IX.  

In the above procedure, care should be taken to allow for the
possibility that the detected particles may be secondary hadrons
arising from  the feeding of heavier primary hadrons.  We shall examine
the fractions of higher meson resonances in Section VIII and Appendix
A.

One works in the center-of-momentum system of the produced hadrons and
arranges the primarily produced hadrons in the order of their
rapidities: $y_{-m}<y_{-m+1}<...<y_{-2}<y_{-1}<\tilde y_0<y_1<y_2<
..<y_{n-1}< y_n$, where $\tilde y_0$ is closest to $y= 0$.
 
The space-time coordinates $(u_i,v_i)$ of the $q$-$\bar q$ production
vertices of the $q\bar q$ pairs for the production of the hadrons in a
single flux tube can be obtained as follows:

\begin{enumerate}

\item
To each hadron with transverse mass $m_{Ti}$ and rapidity $y_i$,
determine $\Delta v_i$ and $\Delta u_i$ by
\begin{eqnarray}
-\frac{\Delta v_i}{\Delta u_i}
=-\frac{ v_{i+1}-v_i}{u_{i+1}-u_i}=e^{-2y_i},\\
\hspace*{-1.2cm}{\rm and~}~~~~~~~~~~~~~~~~~~~~~~~~ \Delta u_i  ~(-\Delta v_i) 
= \frac{m_{Ti}^2}{\kappa^2}.
\end{eqnarray}
The above two equations have the solution
\begin{eqnarray}
(\Delta u_i,\Delta v_i)=\pm  \frac{m_{Ti}}{\kappa}(e^{y_i}, -e^{-y_i}),
\label{eq27}
\end{eqnarray}
where the upper sign on the right-hand side is for positive $y_i$ and
the lower sign for negative $y_i$.

\item
We need to anchor one of the vertices, say the hadron with rapidity
$\tilde y_0 \sim 0$, to a space-time coordinate $(u_0,v_0)$ as
determined from the measured $dN_{pp}/dy$.  In the case of a Gaussian
distribution, they are given by Eqs. (\ref{31}) and (\ref{32}) as a
function of $y$ which is now set to $y=\tilde y_0$, with parameters
$y_0$, $\sigma$, and $m_T$ as determined from other prior measurements
of single-particle rapidity distributions.

\item
Starting with the vertex at $(u_0,v_0)$ as shown in Fig.\ \ref{recon},
we obtain the locations of other pair-production vertices in both
directions of the longitudinal $z$-axis by the recursion formula
\begin{eqnarray}
(u_i,v_i)=(u_{i-1},v_{i-1})+(\Delta u_{i-1},\Delta v_{i-1}).
\end{eqnarray}
 
\vspace*{-0.6cm}
\begin{figure}[h]
\hspace*{0.3cm}
\resizebox{0.45\textwidth}{!}{%
\includegraphics{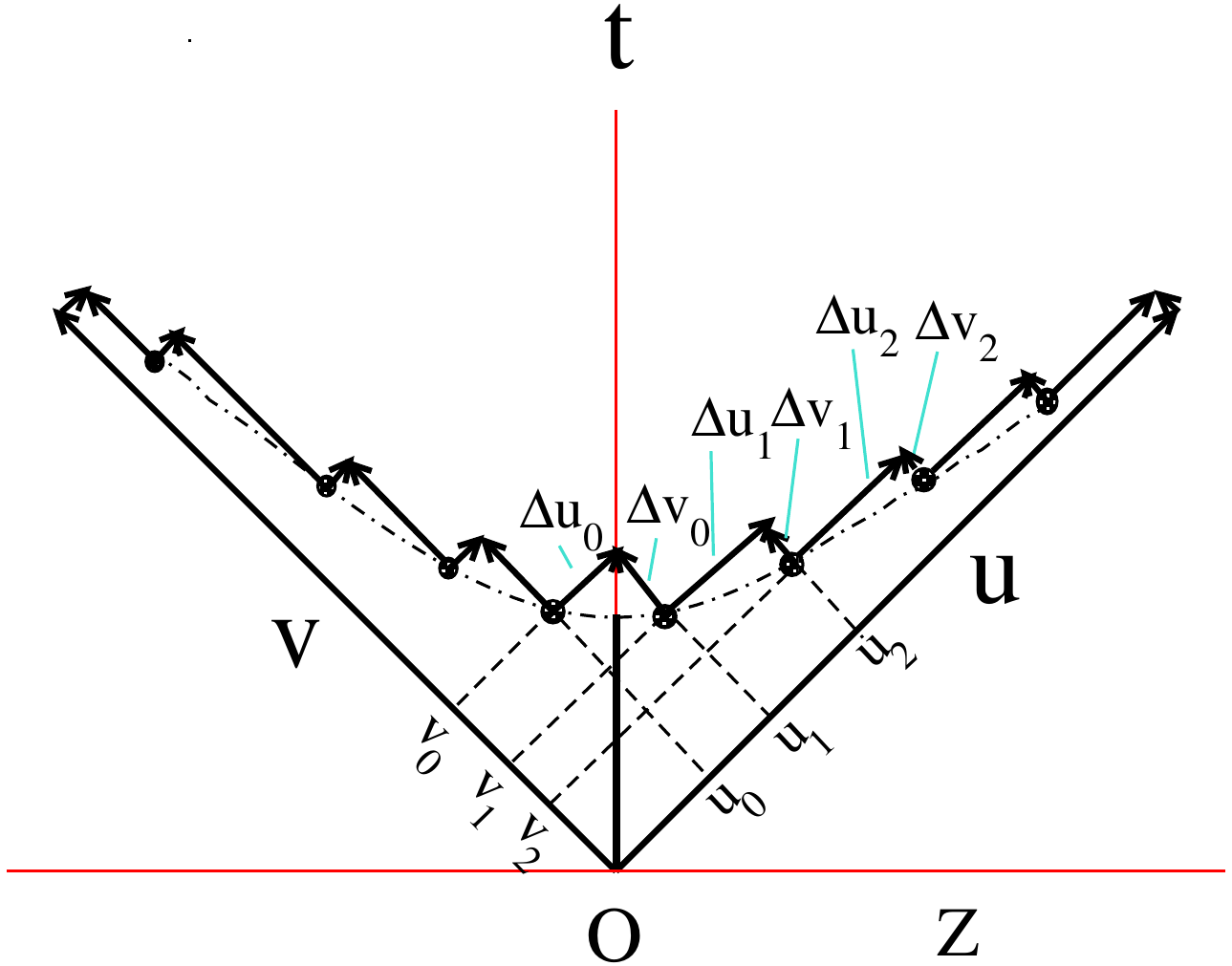}
}
\vspace*{0.2cm}
\caption{ (Color online). Shown here are the forward light-cone
  coordinate $u = z+t$ and backward light-cone coordinate $v=t-z$ for
  the reconstruction of space-time coordinates of vertices $(u_i
  v_i)$, where a $q$ and a $\bar q$ (represented by arrows) are
  produced.  One starts with the vertex located at $(u_0,v_0)$ near
  $\tilde y_0 \sim$ 0 and use the recursion relation,
  $(u_i,v_i)$=$(u_{i-1},v_{i-1}) + (\Delta u_{i-1},\Delta v_{i-1}) $,
  to determine the other pair-production vertices.  }
\label{recon}
\end{figure}

\end{enumerate}

In the above reconstruction process, it should be kept in mind that as
flux-tube fragmentation is only one of the many production mechanisms,
not all collision events arise from flux-tube fragmentation.  It may
be necessary to consider each event case by case to see if it can be
subject to the flux-tube fragmentation analysis.

\section{Experimental Challenges}

There will be experimental challenges in the execution of the
procedures outlined above.  In a collision event with an appropriate
trigger, there will be produced hadrons missing from detection, and
the optimization of the detector will favor the detection of some
hadrons more than some others.  Neutrons, decay photons, and weak
interaction products may also not be recorded.  Nevertheless, with detection
efficiency greater than 90\% for $\pi^-$ \cite{NA61} (and perhaps
slightly lower for other hadrons), there may remain an appreciable
number of exclusive events sufficient for event-by-event
analysis.  One can alternatively lower the goal by examining the fragmentation only in a limited partial section of
the flux tube (e.g. in the forward, backward, or central rapidity region).  The examination
of the many-body dynamics of even a partial section of the flux tube,
no matter how short, remains useful in providing valuable information
on the dynamics of the full flux-tube fragmentation process.

For those exclusive measurements with nearly no missing hadrons, there
remain other complications.  It is general thought that a large
fraction of pions come from the decay of heavier mesons that are
primarily produced in the hadronization process.  The event-by-event
analysis should be carried out with this possibility in mind.
 
The production of heavier mesons can be investigated by evaluating the
invariant masses of combinations of detected lighter observed hadrons.
The complexity of such a procedure depends on the number of produced hadrons
and the 
fraction of heavier
mesons, in comparison to the lighter mesons.  It is therefore useful
to have an idea on the fraction of heavier mesons that may be present
by examining flux-tube fragmentation events with a standard Monte
Carlo event generator.

For such an investigation, we use the the Lund fragmentation model in
the event generator PYTHIA 6.4 \cite{Sjo06}, for hadron production at
the sample energy of $\sqrt{s}$=17.3 GeV in the NA61/SHINE energy
range, corresponding to a fixed target $pp$ collision at $p_{\rm
  lab}$=158 GeV/c.  
In the default PYTHIA 6.4 program, the vector meson fraction involving
$u$ and $d$ quarks is
$(V/(V+P))_{ud}$$\equiv $$f_V^{ud}$=PARJ(11)=0.5, the vector meson
fraction involving the $s$ quark is $(V/(V+P))_{s}$$\equiv$$f_V^{s}$=
PARJ(12)=0.6, and the strangeness suppression factor is
$(s/u)$$\equiv$$ f_s$=PARJ(2)=0.3. 
One finds that a very large fraction of produced
primary hadrons are $\rho$, $\omega$, $K$ and $K^*$ mesons.

 Upon a careful examination and comparing with experimental
 measurement of heavier meson production, one finds that the primary vector and strange meson
 fractions may be over-estimated in the NA61/SHINE and SPS
 energy ranges  in PYTHIA 6.4 with the default
 parameters, and a fine tuning appropriate for the range of energies
 of our interest may be needed.  Previously in the production of
 $\rho^0$, $\omega$ and $K^*$ by $e^+ e^-$ annihilation in the
 $\Upsilon$ energy region at $\sqrt{s}({e^+e^-})$=9.46 to 10.02 GeV, a
 similar observation of an over-estimation with the default parameters was
 reached \cite{Alb94}, using the Lund 7.3 Monte Carlo program that was
 adopted as part of the Lund fragmentation scheme in PYTHIA 6.4.

 What then is the primary heavy meson fraction for 
the case of $pp$ collisions at $\sqrt{s}\sim $ 10 to 20  GeV at which flux tube fragmentation is expected to dominate?
At  $\sqrt{s}=$ 17.3 GeV,  the 
experimental observed heavier meson  ratios  are $\sigma_{\rm
  obs}(\rho^0)/\sigma_{\rm obs}(\pi^-)$=0.12 \cite{Sin76,Sin78},
$\sigma_{\rm obs}(K^-)/\sigma_{\rm obs}(\pi^-)$ = 0.08 \cite{Pul15},
and $\sigma_{\rm obs}(K^{*+}) / \sigma_{\rm obs}(K^+)$=0.32
\cite{Sin78}.  As discussed in Appendix A, these observed ratios give
the slightly modified primary ratios to be $\sigma_{\rm pri}(\rho^0)/\sigma_{\rm
  pri}(\pi^-)$=0.19, $\sigma_{\rm pri}(K^-)/\sigma_{\rm
  pri}(\pi^-)$=0.09, and $\sigma_{\rm pri}(K^{*+})/\sigma_{\rm
  pri}(K^+)$=0.47, after taking into account the feeding of heavier
mesons.  Using these primary ratios for the energy range of our
interest, we find in Appendix A the heavy meson fractions
$f_V^{ud}$=$0.20$, $f_s$=$0.14$, and $f_V^{s}$=$0.32$.  These coefficients leads to   dominant primary heavier
meson fraction $f_V^{ud}$  only at the 20\% level.   With a multiplicity of order 10-15 in
each event,  the identification of the primary higher mass mesons at the NA61/SHINE and NICA energies may perhaps be a manageable task.

\section{Separation of the two flux tubes in a $pp$ collision}

At NA61/SHINE collision energies,  Ref. \cite{NA61} gives  the rapidity
distribution $dN(\pi^-)/dy$ in Fig. 2(b) as the sum of
two Gaussian distributions with rapidity displacements at $y_0$ and
$-y_0$.  
The two displaced Gaussian rapidity distributions can be 
considered  as an indirect evidence for the presence of two flux tubes 
in $pp$ collisions.   An event-by-event analysis  may provide a way to
discriminate the the possibilities of one,  two, or many  flux tubes.  

We enlist  below useful tools that may be utilized in the 
reconstruction of the flux tubes:

(1) Because of the conservation of baryon numbers, there will be two
baryons in each $pp$ flux-tube fragmentation event.  As leading particles, these baryon
products can be used to anchor the two ends of the two flux tubes at
opposite rapidities.  Because
neutrons are difficult to detect, one can choose events with two
charged baryons to be the anchors at the opposite ends of the two flux
tubes, and look for hadrons linking two separate tubes.

 (2) Because of charge conservation in $q$-$\bar q$ pair production, a
hadron adjacent to a charged hadron from the same tube is unlikely to
have a charge of the same sign. 

(3) To every produced strange hadron, there is an adjacent hadron of
opposite strangeness from the same tube, on account of flavor
conservation in $q$-$\bar q$ pair production.  The flavor quantum
numbers of the chain of hadrons from a tube are aligned.   The flavor of the quark in a  hadron and the flavor of the antiquark of the hadron adjacent in rapidity are the same in the semi-classical picture.  
Such a correlation may be modified by quantum and other fluctuation mechanism, and experimental
measurements of such correlations are sorely needed for a quantitative assessment.  

(4) In the flux-tube fragmentation under the pair production mechanism,
a hadrons adjacent to each other in the flux tube is likely to be
correlated azimuthally at large opening angles close to $\pi$ because the produced
$q$ and $\bar q$ are produced back-to-back azimuthally in the pair
frame.   As
the transverse momenta of the other spectator quark and antiquark are not
necessarily zero, the back-to-back azimuthal correlation is reduced to
be a correlation at a large opening azimuthal angle close to $\pi$.  

(5) If the rapidity centers of hadrons produced by two flux tubes are
displaced by $-y_0$ and $+y_0$, as given in Eq.\ (\ref{34}) by
\cite{NA61}, then the two flux tubes will produced hadrons that are
displaced from each other in rapidity.   The rapidity displacement $y_0$ increases with
the collision energy, as indicated in Table II.  As a consequence, the
sections of hadrons of the two flux tubes that overlap in rapidity is
smaller, the greater the displacement of the two flux tube rapidities
$y_0$ and the easier it is to separate the two flux tubes.  On the
other hand, the greater the collision energy, the greater the
multiplicity and the more difficult  it is to reconstruct the dynamics with a
greater number of possible combinations.

With this set of tools, it may be possible to resolve some of the
puzzles of the hadron configuration for some of the $pp$ collision
events.  Because no such data have been developed as yet, the
knowledge of some explicit space-time dynamics in some events serves
well in extending our understanding of the hadronization dynamics of the
particle production process.

\section{Inside-Outside Cascade or Outside-Inside Cascade?}

An explicit reconstruction of the many-body correlations provides a
valuable tool to distinguish the fine details of different reaction
mechanisms of flux-tube fragmentation  in high-energy $e^+e^-$ annihilations and $pp$ collisions.

There is the outside-inside cascade picture of particle production
in flux-tube fragmentation
that comes from perturbative QCD in which the outgoing quark and
antiquark (or diquark) are represented by longitudinal jets of
partons, and hadrons are produced by parton cascading from the jets.
Low energy hadrons arises from the ``wee" partons which could form a
bridge and neutralize the quark color charges with those from the
cascading of the other jet.  This is the outside-inside cascade
picture of the longitudinal production of soft particles.

An alternative view that has been suggested by Bjorken, Casher,
Kogut, and Suskind \cite{Bjo73,Cas74} is the inside-outside
longitudinal cascade in which the receding quark and antiquark pair
interact by the exchange of a vector interaction (gluon exchange), and
these interactions lead to the production of particles from the inside
out.

The explicit display of the dynamics as a result of the flux-tube
fragmentation will allow an assessment of the mechanism of particle
production.  The difference between the inside-outside and
outside-inside picture shows up as a difference in the correlation of
the soft particles, those with small rapidities.  In the
inside-outside cascade picture, the hadrons in the small rapidity
region are correlated because they share the process of
nonperturbative particle production, while in the outside-inside
cascade, these particles are not correlated as they come from
independent cascade of the leading partons.  At the time when the soft
particles are produced, the leading partons are so far apart that
there is little chance for these wee partons to overlap in coordinate
space to correlate.

The two-particle angular correlation with opposite charges may appear
to lend support to the picture of inside-outside cascade for flux-tube
fragmentation for low-$p_T$ particle production \cite{Won15}.  The
situation may depend on the transverse momentum $p_T$ of the produced
particles, as the hard process with high-$p_T$ particles may favor the
outside-inside cascade picture of particle production.  A many-body
correlation will be able to reveal whether the picture of
inside-outside or outside-inside cascade will depend on the $p_T$
domain of the detected particles.

\section{Conclusion and Discussions}

There are many mechanisms of particle production in $pp$ collisions,
such as flux-tube fragmentation
\cite{Nam70,Bjo73,Cas74,Sch62,Art74,And79,And83,Art84,And83a,Sjo86,Sjo06,Sjo14,Sch51,Wan88,Pav91,Won91a,Won91b,Gat92,Won95,Feo08,Won94},
hard-scattering \cite{Bla74,Ang78,Fey78,Owe78,Rak13,Eli96,Won12},
direct fragmentation \cite{Won80}, color-glass condensate
\cite{Mc94,Mc01}, EPOS model \cite{Wer06}, and Landau hydrodynamics
\cite{Lan53,Won08,Mur04,Ste05,Sen15,Won14}.  Each of the proposed
theoretical production mechanisms exhibits specific space-time
dynamics on an event-by-event basis.  If a substantial fraction of the
produced particles are detected in the collision event, then the
event-by-event dynamics of the produced particles may be utilized to
identify the particle production mechanism that has occurred in that
event, not only to test and discriminate theoretical descriptions but
also to provide important pieces of information on the production
process.  It is therefore useful to develop methods to analyze
event-by-event dynamics of produced hadrons in order to extract the
wealth of information they provide.
For clarity, focus of physical principles, and a better theoretical
understanding, we have examined the flux-tube fragmentation process
with analytical models and simplifying assumptions
against which refinement and modifications  may be brought forth
for a comprehensive description of the production and hadronization process.

Successes in carry out the event-by-event analysis in flux-tube
fragmentation may stimulate similar exploration to investigate the
space-time dynamics in other reaction mechanisms such as the
space-time dynamics in a hard-scattering process in which  hadrons from azimuthally back-to-back jets are expected and
the dynamics of the showering process are displayed.  Similarly, a
hydrodynamical occurrence in a collision event with a large number of
produced particles will exhibit specific patterns in an event-by-event
basis \cite{Sen15,Won14}.  The simple
event-by-event analysis in flux-tube fragmentation 
 in $pp$
may assist  the
investigation of the partition of the two flux tubes 
collisions, the interference and interaction between flux tubes, the
possible merging of the flux tubes, and the description of the multiple
collision processes in $pA$ and $AA$ collisions.

In $pp$ collisions at $\sqrt{s}= $6-200 GeV, the gross features
of the experimental NA61/SHINE \cite{Mak15,Gaz15,Lar15,Ser15} and STAR
\cite{STAR06twopar,Por05,Tra11,Ray11} angular correlation data for two
low-$p_T$ hadrons with opposite charges match the signature of the
fragmentation of a flux tube \cite{Won15}.  Thus $pp$ collisions in
the energy range of the NICA facility and NA61/SHINE energy scan are
favorable for the study of the many-body correlation for flux-tube
fragmentation discussed here.  
 
Exclusive measurements has the difficulties that there will be
particles missing from detection, and the optimization of the detector
will favor the detection of some hadrons more than some others.  It is
an experimental question whether there remains an appreciable
fractions of events in which all (or almost all) produced hadrons may be 
recorded.  If this is difficult to achieve, a lower goal of
looking only for the fragmentation process in a partial section of the
whole flux tube in the forward, backward, or central rapidity region,  may  be carried out with
the procedures outlined in Section VII.    The ability to examine the
many-body dynamics of even a partial section of the flux tube may be a useful
asset.
 It will pave the way
for a full investigation of the full flux tube when detection
conditions may become more favorable. 

With regard to the rapidity-space-time ordering that is expected
semi-classically, there has not been much study on the quantum
fluctuation of such an ordering.  The explicit event-by-event
measurement will provide useful information on how the flux-tube
fragmentation will be affected by stochastic and quantum fluctuations.

If confirmed experimentally in $pp$ collisions, the
rapidity-space-time ordering may be used as a strong signature of
flux-tube fragmentation that may be applied to the search for the
possible presence of the remnants of cosmic flux tubes in the early
history of the universe.  One can envisage that at the end-point of
the expansion phase of the cosmic string dynamics, the cosmic strings
will fragment into cosmic mass entities.  If the analogy of color
string fragmentation is a good guide, the longitudinal velocities or
rapidities of cosmic mass entities will be ordered longitudinally
along the string axis and adjacent cosmic masses may be correlated
azimuthally with back-to-back transverse momenta.  A measurement of
the angular correlation between cosmic masses and the momenta of a
longitudinal chain of masses may provide a glimpse of the space-time
dynamics of the prior flux tube configuration and a test of the
concept of knotty/linked flux tubes proposed recently \cite{Ber15}.

\vspace*{0.3cm}
\centerline{\bf Acknowledgements}
\vspace*{0.3cm}

The author would like to thank Drs. David Blaschke, Elena Kokoulina,
and Ken Read for helpful discussions and to Prof.\ A.\ Berera for
bringing attention to the knotty/linked flux tubes in cosmological
inflation.  The research was supported in part by the Division of
Nuclear Physics, U.S. Department of Energy under Contract
DE-AC05-00OR22725.

\vspace*{-0.6cm}
\appendix 

\section{Vector meson fractions and the Strangeness suppression factor  } 

Experimental measurements in $pp$ collisions and $e^+$-$e^-$
annihilations give ratios of observed yields $\sigma_{\rm
  obs}(\rho^0)/\sigma_{\rm obs}(\pi^-)$, $\sigma_{\rm
  obs}(K^-)/\sigma_{\rm obs}(\pi^-)$, and $\sigma_{\rm
  obs}(K^{*-})/\sigma_{\rm obs}(K^-)$ where the subscript ``obs"
stands for observed hadron quantities, to distinguished from the
primary hadron quantities with the subscript ``pri".  For $pp$
collisions involving flux-tube fragmentation, the ratios of neutral
or negative hadrons are less likely to arise from leading projectile
and target fragmentations and are more likely to come from flux-tube
fragmentation.  We would like to examine those yield  ratios involving
neutral and negative hadrons within the mechanism of flux-tube
fragmentation.

To infer the primary yield  ratios  within the flux-tube
fragmentation mechanism, we need to take into account the feeding from
the heavier mesons in $\rho\to \pi \pi$, $\omega \to \pi \pi \pi$ and
$K^* \to (K \pi)$.  The observed and the
primary cross sections are related by
\begin{eqnarray}
&&\sigma_{\rm obs} (\pi^+ +\pi^0 + \pi^-)=\sigma_{\rm pri} (\pi^++\pi^0+\pi^-)
\nonumber\\
&&\hspace*{2.5cm}+2 \sigma_{\rm pri} (\rho^++\rho^- +\rho^0)
+3 \sigma_{\rm pri}(\omega)
\nonumber\\
&&\hspace*{2.5cm}
+ \sigma_{\rm pri} (K^{*+}+K^{*-}+K^{*0}+{\bar K}^{*0}),
\nonumber\\
&&\sigma_{\rm obs} (K^+\!+K^-\!+K^0\!+\bar K^0)=\sigma_{\rm pri} (K^+\!+K^-\!+K^0\!+\bar K^0)
\nonumber\\
&&\hspace*{2.5cm}
+\sigma_{\rm pri} (K^{*+}+K^{*-}+K^{*0}+{\bar K}^{*0}).
\nonumber
\end{eqnarray}
 The ratios of the primary yields can then be used
to determine the vector meson fractions $f_V^{ud}$ and $f_V^s$, and
the strangeness suppression factor $f_s$.  They are  defined
as
\begin{eqnarray}
 f_V^{ud}&&= \left ( \frac{V}{V+P} \right )_{\!\!\!ud} 
\nonumber\\
&&=
\frac{\sigma_{\rm pri} (\rho^++\rho^-+\rho^0+\omega)}
{\sigma_{\rm pri} (\pi^++\pi^0+\pi^-\!+\rho^++\rho^-+\rho^0+\omega )},
\nonumber\\
f_V^{s}&&= \left ( \frac{V}{V+P} \right )_{\!\!\!s} 
\nonumber\\
&&=
\frac{\sigma_{\rm pri} (K^{*+}\!+K^{*-}\!+K^{*0}\!+{\bar K}^{*0})}
{\sigma_{\rm pri} (K^+\!\!+K^-\!\!+K^0\!+\bar K^0\!+K^{*+}\!\!+K^{*-}\!\!+K^{*0}\!\!+{\bar K}^{*0})},
\nonumber\\
\!f_s&&\!=\!\!\frac{\sigma_{\rm pri} (K^+\!\!+K^-\!\!+K^0\!+\bar K^0\!+K^{*+}\!\!+K^{*-}\!\!+K^{*0}\!\!+{\bar K}^{*0})}
{\sigma_{\rm pri} (\pi^+\!+\pi^0\!+\pi^-\!+\rho^+\!+\rho^-\!+\rho^0\!+\omega)}.
\nonumber
\end{eqnarray}
We assume that in a $pp$ collision, the fragmentation of the leading
charged hadrons have been taking into account, and we are left to deal
with a neutral quark-antiquark color flux tube (or tubes), for which
$\sigma_{\rm pri} (\rho^+)$=$\sigma_{\rm pri} (\rho^0)$=$\sigma_{\rm
  pri} (\rho^-)$=$\sigma_{\rm pri} (\omega)$, $\sigma_{\rm pri}
(K^{\pm })$=$\sigma_{\rm pri} (K^0)$=$\sigma_{\rm pri} (\bar K^0)$,
and $\sigma_{\rm pri} (K^{*\pm })$= $\sigma_{\rm pri}
(K^{*0})$=$\sigma_{\rm pri} (\bar K^{*0})$.  After taking into account
the feeding  from the heavier mesons to of lighter mesons, we obtain for the
fragmentation of a neutral color flux tube
\begin{eqnarray}
\frac{\sigma_{\rm pri}(\rho^0)}{\sigma_{\rm pri}(\pi^-)} &&= \frac{{\sigma_{\rm obs}(\rho^0)/}{\sigma_{\rm obs}(\pi^-)}}{1-3{\sigma_{\rm obs}(\rho^0)/}{\sigma_{\rm obs}(\pi^-)}}
\nonumber\\
&&\times \left (1+ \frac{f_V^{s}}{1-f_V^{s}}\frac{4}{3}\frac{\sigma_{\rm pri}(K^-)}{\sigma_{\rm pri}(\pi^-)}\right ),
\end{eqnarray}
\begin{eqnarray}
\frac{\sigma_{\rm pri}(K^{-})}{\sigma_{\rm pri} (\pi^-)}
=\frac{
\left ( 1+
\frac{9}{4} \frac{f_V^{ud}} {(1-f_V^{ud})}   \right ) 
\frac{\sigma_{\rm obs}( K^{-})}{ \sigma_{\rm obs} (\pi^-)}   } 
{ \frac{1}{1-f_V^s} -\frac{4}{3}  \frac{f_V^s}{1-f_V^s}\frac{\sigma_{\rm obs}( K^{-})}{ \sigma_{\rm obs} (\pi^-)} 
},
\end{eqnarray}
\begin{eqnarray}
\frac{\sigma_{\rm pri}(K^{*-})}{\sigma_{\rm pri} (K^-)}=
\frac{\frac{\sigma_{\rm obs}(K^{*-})} {\sigma_{\rm obs}(K^-)}}
{1-\frac{\sigma_{\rm obs}(K^{*-})} {\sigma_{\rm obs}(K^-)}},
\end{eqnarray}
where
\begin{eqnarray}
f_{V}^{ud}=\left ( \frac{V}{V+P}\right )_{ud} = \frac{\frac{4}{3}\frac{\sigma_{\rm pri}(\rho^0)}
{\sigma_{\rm pri}(\pi^-)}}
{1+\frac{4}{3}\frac{\sigma_{\rm pri}(\rho^0)}{\sigma_{\rm pri}(\pi^-)}},
\end{eqnarray}
\begin{eqnarray}
f_V^s
= \frac{ \frac{\sigma_{\rm pri}(K^{*-})} {\sigma_{\rm pri}(K^-)}}
{1+ \frac{\sigma_{\rm pri}(K^{*-})} {\sigma_{\rm pri}(K^-)}},
\end{eqnarray}
\begin{eqnarray}
f_s=\frac{1-f_V^{ud}}{1-f_V^s} \frac{4}{3} \frac{\sigma_{\rm pri}(K^-)}{\sigma_{\rm pri}(\pi^-)}.
\end{eqnarray}
This set of equations allow the determination of the primary vector meson
fraction and strangeness suppression factor from the observed yield
ratios in flux-tube fragmentation by iteration.

The $\rho^0$ and $\pi^-$ meson yields in $pp$ collisions at NA61/SHINE
and SPS energies have been measured experimentally
\cite{Blo74,Alb79,Sin76,Sin78}.  At $\sqrt{s}$=6.84 GeV, the observed
$\sigma_{\rm obs}(\rho^0)/\sigma_{\rm obs}(\pi^-)$ has been found to
be about 0.07 \cite{Blo74}.  At $\sqrt{s}$$\sim$ 19.7 GeV, it rises to
0.12$\pm$0.3, but corrections to the background $K$-$\pi$ correlations
in $\pi$-$\pi$ correlation measurements reduce this observed
$\rho^0/\pi^-$ ratio to 0.09$\pm$0.03 \cite{Sin76,Sin78}.  At higher
energies up to $\sqrt{s}$$\sim$ 60 GeV, the ratio $\sigma_{\rm obs}(\rho^0)/\sigma_{\rm obs}(\pi^-)$  becomes nearly flat
at about 0.13 (See Fig. 5 of Ref. \cite{Alb79}).  These ratios are
consistent with the multiplicity ratio $\langle \rho^0\rangle/ \langle
\pi^-\rangle$ in $e^+$-$e^-$ annihilation, which gives $\langle
\rho^0\rangle/ \langle \pi^-\rangle $= $\{0.109, 0.138, 0.138\}$ at
$\sqrt{s}(e^+e^-)=$$\{ 10, 25$-$35, 91\}$ GeV, respectively
\cite{PDG14}.

Experimental measurements in \cite{Sin78} give an observed
$\sigma_{\rm obs} (K^{*-})/\sigma_{\rm obs} (K^{0})$=0.32 (error +0.14, -0.18) and a
strange vector meson fraction $f_V^{s}$=0.32 for $pp$ collisions at
$\sqrt{s}$=19.7 GeV.  Such a ratio is consistent with the multiplicity
ratio in $e^+$-$e^-$ annihilations, which give observed $\langle K^{*+}\rangle/
\langle K^+\rangle$=(0.29$\pm$0.04, 0.43$\pm$0.04) at $\sqrt{s}({e^+e^-})$= (10.45,
29) GeV, respectively (Table 18 of  \cite{Alb94}).

In $pp$ collisions at NA61/SHINE energies, the ratio $K^+/\pi^+$ is
greater than $K^-/\pi^-$.  At these low energies the production of
$K^+$ and $\pi^+$ is influenced by contributions from baryon resonance
production in which an incident proton is excited to a baryon
resonance that subsequently emits a meson \cite{Gaz15,Lar15}.  The
baryon resonances are produced mostly at the projectile and target
fragmentation regions but have contributions at central rapidity when
the collision energies are low.  On the other hand, negative mesons
such as $K^-$ and $\pi^-$ are more likely to arise from flux-tube
fragmentation rather than projectile or target fragmentation.  The
ratio $K^-/\pi^-$ can be a better gauge of the strangeness suppression
in flux-tube fragmentation.

At $\sqrt{s}$$\sim$3 GeV, the ratio $K^-/\pi^-$ is zero because the
threshold for $K^-$ production is higher than the threshold for
$\pi^-$ production.  At $\sqrt{s}$=17.3 GeV, $K^-/\pi^-$ rises to
about 0.08.  At $\sqrt{s}$=200 GeV, one finds $\sigma_{\rm
  obs}(K^+)$$\sim$$ \sigma_{\rm obs}(K^-)$ and $\sigma_{\rm
  obs}(\pi^+)$$\sim$$ \sigma_{\rm obs}(\pi^-)$, with $\sigma_{\rm
  obs}(K^+)/ \sigma_{\rm obs}(\pi^+)$$\sim$$\sigma_{\rm obs}(K^-)/
\sigma_{\rm obs}(\pi^-)$$\sim$0.1 \cite{Ada04}.  At higher energies,
the ratio $\sigma_{\rm obs}(K^-)/ \sigma_{\rm obs}(\pi^-)$ become
relatively flat at 0.12 (see Fig. 7 of \cite{Pul15}).

As an example, we can consider~the~case~of~$\sqrt{s}$=17.3 GeV
at which  $\sigma_{\rm obs}(\rho^0)/\sigma_{\rm obs}(\pi^-)$=0.12\cite{Sin76,Sin78}, $\sigma_{\rm obs}(K^{*+}) / \sigma_{\rm
  obs}(K^+)$=0.32 \cite{Sin78}, and $\sigma_{\rm obs}(K^-)/\sigma_{\rm
  obs}(\pi^-)$=0.08 \cite{Pul15}.  We find from~Eqs.\ (A1)-(A3) the ratios of primary
yields to be $\sigma_{\rm pri}(\rho^0)/\sigma_{\rm pri}(\pi^-)$=0.19,
$\sigma_{\rm pri}(K^-)/\sigma_{\rm pri}(\pi^-)$=0.09 \cite{Pul15}, and
$\sigma_{\rm pri}(K^{*+})/\sigma_{\rm pri}(K^+)$=0.47.  From Eqs.\ (A4)-(A6), 
these primary
ratios give $f_V^{ud}$=0.20, $f_V^{s}$=0.32, and $f_s$=0.14.

\end{document}